\documentclass[runningheads,citeauthoryear]{apinv}
\usepackage{epsfig,cite,graphics}
\usepackage{xcolor}
\usepackage{marvosym} %For astronomical symbols % % %
\usepackage[utf8]{inputenc}

\begin{document}

\title{A New Approach to Find the B{\"o}hm-Vitense gap}
\titlerunning{}
\author{Tahereh Ramezani\inst{1}, Ernst Paunzen\inst{1}, Martin Piecka\inst{2}, and Michal Kajan\inst{1}}
\authorrunning{T.Ramezani, E.Paunzen, M.Piecka, and M.Kajan}
\tocauthor{Tahereh Ramezani, Ernst Paunzen, Martin Piecka, and Michal Kajan} 
% Command tocautor{} is used by the Latex to give author names 
% to the Contents of the volume (automatically generated)
\institute{Department of Theoretical Physics and Astrophysics, Masaryk University, Kotl\'a\v{r}sk\'a 2, CZ-611\,37 Brno, Czechia
	\and Department of Astrophysics, Vienna University, T{\"u}rkenschanzstraße 17, A-1180 Vienna, Austria
	  \newline
	\email{514009@mail.muni.cz}    }
\papertype{Submitted on 13.11.2023; Accepted on 05.01.2024}	

% Papertype can be "Research report", "Review", "Invited lecture", "Conference talk", 
% "Conference poster", "Lecture at scientific seminar", "Summary of dissertation",  etc.
\maketitle
\pagenumbering{gobble}

\begin{abstract}
This paper discusses the B{\"o}hm-Vitense gap, a gap in the colours of stars that occurs when the atmosphere changes from radiative to convective in deep layers. We are using different algorithms for detecting gaps in colour-magnitude diagrams (CMDs), including the k-nearest neighbours (k-NN) and UniDip algorithms. We propose using a combination of the k-NN algorithm and the UniDip algorithm and manual verification to identify gaps unlikely to be of a statistical origin. Using the $Gaia$ photometric system, i.e. $BP-RP$, we took the data of 130 star clusters and searched for
gaps in the ranges of 0.40 to 0.47\,mag, and 0.56 to 0.60\,mag, respectively.  We analysed all data statistically and identified the gaps in the individual clusters. Finally, we applied the kernel density estimator to see how the gaps are distributed.
\end{abstract}

\section*{Introduction}
B{\"o}hm-Vitense, (1970) was the first to realise that there could occur a gap in the colours of stars when the atmosphere changes from radiative to convective in deep layers, at $T_\mathrm{eff} \approx $ 7500 K [or at $ 0.22 \leq  B - V \leq  0.28 $ mag;] B{\"o}hm-Vitense \& Canterna, (1974).

There are two possible ways in which convection might alter the colour: either because (1) the temperature gradient in deep atmospheric layers becomes smaller than the radiative gradient, and the lower temperatures provide a relatively more minor flux in the $B$ and $U$ bands, leading to a sudden slight increase in the $B-V$; or (2) the photospheric granulation associated with convection originates temperature inhomogeneities, which again redden the $B-V$. Observations of this gap in the field population have generally been obtained using spectroscopically determined luminosity classes to select only main-sequence (luminosity class V) objects. For field stars, the gap is typically in the $0.2 < B - V < 0.3$ range, which is consistent with the prediction of B{\"o}hm-Vitense, (1970)(B{\"o}hm-Vitense \& Canterna, (1974)). 

One would imagine that such a gap, as evident as it is in the field population, would be easily detected in open clusters.
Observations of clusters, however, have produced ambiguous results. For example, Mendoza \& Eugenio, (1956) finds no gaps in the Pleiades, while Jasniewicz, (1984), using stricter selection criteria, does note a gap. The latter author observes the position in the colour of the blue edge of the B{\"o}hm-Vitense gap to differ with cluster age. Canterna et al., (1979) and Kjeldsen \& Frandsen, (1991) note that there are many different gaps, while Harris et al., (1993) find hints of a gap but conclude that it may not be significant. Even B{\"o}hm-Vitense \& Canterna, (1974) do not see a gap in all clusters. The statistical significance, location in colour space, and density depression of the gaps appear to change from cluster to cluster.

One might account for the differing positions of the gaps in different clusters if the rapid rotation of the stars in some clusters systematically retarded the onset of convection B{\"o}hm-Vitense \& Canterna, (1974).

B{\"o}hm-Vitense, (1982) suggested that differing rotations, and thus different critical temperatures for the onset of convection, would cause a spread in a plot of $T_\mathrm{eff}$ versus $B-V$. Based on a sample of about 50 stars, her paper finds evidence for two branches of late-A and early-F stars.
Here $T_\mathrm{eff}$ was determined from ultraviolet observations that probe higher atmospheric layers, which were claimed by B{\"o}hm-Vitense, (1982) to be a better predictor of a star’s $T_\mathrm{eff}$ than the $B-V$ colour index. Recently, however, Simon \& Landsman, (1997) obtained more UV observations of A and F stars and found no evidence for a bifurcation in the ($B-V$, $T_\mathrm{eff}$) plane; they suggest that there may be no such phenomenon.   

Convective energy transport in the atmosphere of a star will redden the star because of the lower temperatures in deeper layers, which contribute to the surface flux in the spectral region of the U and B filters. Since the mixing-length theory of convection predicts an abrupt onset of convection at a given temperature $T1$, all stars with
$T_\mathrm{eff} > T1 $ will be in radiative equilibrium, while all those with $T_\mathrm{eff} < T1$  will have partly convective energy transport. If a radiative star with $T_\mathrm{eff}$ = $T1$ has a colour $B - V = C$, then a star with $T_\mathrm{eff}$ = T1 - $\epsilon$ ($\epsilon$ being a very small number) will have $B - V = C + \Delta$, where $\Delta$ is the change of colour due to convection. If the onset of convection is abrupt, then there should be no stars with $C < B-V < C + \Delta$; i.e., we should expect a gap in $B-V$. The size of the gap is determined by the parameter A, which depends on the amount of convective energy transport at the onset of convection. If a scaled solar temperature stratification were assumed for the convective star, we would derive $\Delta$ = 0.07 if the microturbulence were the same in the radiative and convective stars B{\"o}hm-Vitense \& Canterna, (1974). If microturbulence is larger for convective stars, then A should be larger. If convection were less effective in F stars than in the Sun, then A should be smaller.
The gap in B-V should occur at that $B-V$ for which $T_\mathrm{eff} = T1$. Therefore, the position of the gap, if it exists, tells us the $T_\mathrm{eff}$ where convection starts to be important for energy transport. \footnote{{The General Catalogue of Photometric Data: http://gcpd.physics.muni.cz/}}

%The B{\"o}hm-Vitense gaps are associated with sudden changes in the structure of convective atmospheres. B{\"o}hm-Vitense, (1982)

%There are two explanations for the colour change. First, the temperature gradient in deep atmospheric layers becomes smaller than the radiative gradient, creating a relatively more minor flux at a higher wavelength. Secondly, temperature inhomogeneities are caused by photospheric granulation, which originates in the convective zone. 

The "original" gap is interpreted as a colour gap. However, some gaps can be interpreted as $T_{{eff}}$ gaps, as shown using the full spectrum theory model (FST) see (D'Antona et al., (2002)).
The problem with calculating gaps in the mixing length theory (MLT) 1D models, is that we use the free parameter $\alpha$ - mixing length parameter, representing a highly simplified convection effectiveness as the length of the mean free path, which convective material will travel. 
It is possible to derive $\alpha$ parameters from a 3D model, but it shows that it is dependent on metalicity, effective temperature, and surface gravity (Magic et al., (2015)).

Sonoi et al., (2019) have shown that FST and MLT 1D models do not entirely correlate with 3D convection models.

\section*{2. Where to look for the gap?}
To look for the gap in individual open B{\"o}hm-Vitense clusters, several criteria must be satisfied beyond the fact that the late-A and early-F stars must lie on the main sequence. First, the cluster must be well surveyed to a point on the main sequence considerably redward of the expected gap. Second, the cluster must be rich enough to obtain reasonable statistics. Third, the photometry must have high internal consistency. Fourth, the photometry must be accurate. Fifth, the colour excess for the cluster must be accurately known, and if the colour excess is variable across the cluster, this must be known accurately Rachford \& Canterna, (2000).

People have used different methods to find the gaps. For example, D'Antona and her colleagues used the Vienna University model atmospheres computed with either Mixing length Theory (MLT) or Full Spectrum of Turbulence (FST) as boundary conditions to interior computation adopting the same convection model.
They find that the models with FST provide a discontinuity in the mass versus colour relation at $B-V <$ 0.4. The MLT models do not show this behaviour at all. Simulations of the colour distributions show that this feature may produce a gap at 
$B-V <$ 0.4 and a clustering of stars at 0.4 $< B-V <$ 0.5. They obtain a new and convincing explanation for the reality and location of this gap. It must be a $T_\mathrm{eff}$ gap owing to the sudden onset of convection in the stellar envelope, and as such, it must also be found in other colours, as confirmed by the analysis of the $V-I$ colour distribution in
the Hyades.
As the B{\"o}hm-Vitense gap can be related to the transition between “radiative” and “convective” envelopes, the FST convection can be used as a tool to test these structures since it joins the observationally necessary very low efficiency in the surface regions and the very high efficiency in the
deeper stellar regions.
The stellar structures computed with FST showed a very sharp transition from
``radiative'' to ``convective'' envelopes D'Antona et al., (2002).

\section*{3. Detecting gaps in a CMD}
In this section, we summarise the methods used by various authors to detect gaps in colour-magnitude diagrams. Furthermore, we also introduce a couple of other ways to approach this problem.

It is well known that the data distribution in the colour-magnitude diagram is non-uniform and varies as a function of the extinction, the distance and the age. This distribution will affect any attempt to search for a gap. It should also be noted that the expected size of the B{\"o}hm-Vitense gap is $\leq 0.10$~mag in $B-V$ Rachford \& Canterna, (2000).

Finally, we would like to point out that our main aim is to search for various gaps in a statistical sense. Some methods applied in the literature can be beneficial but can also be challenging to implement in a semi-automatic and statistical approach.

\subsection*{3.1. Empirical cumulative distribution function}

Aizenman et al., (1969) studied the colour-magnitude diagrams of two old open clusters, NGC~188 and NGC~2682. They aimed to determine the helium content $Y$ of the cluster members using a gap located near the main-sequence turnoff point. This can be done using a relation between $Y$ and the gap size.

The authors used and analysed a variation (sometimes called the Aizenman, Demarque, and Mermilliod (ADM) plot) of the photometric data's empirical cumulative distribution function (CDF) to detect the gap. While such an approach is quite robust, it is sensitive to the number of data points in the diagram. Since we are interested in searching for more than just the B{\"o}hm-Vitense gap, this method might detect gaps which are not real. This can happen if the number of the analysed cluster members is relatively small.

Furthermore, the B{\"o}hm-Vitense gap can remain undetected in the cumulative distribution function while being present in the CMD when viewing it by eye. Such a situation is most likely to occur when a significant number of data points is located in the gap due to the presence of binary and peculiar stars de Bruijne et al., (2001).

\subsection*{3.2. Histogram analysis}

Kovtyukh et al., (2004) note the presence of a previously unidentified gap at $\sim 0.70$ mag. Their study focused on analysing the effective temperature distribution of a sample of nearby field dwarfs. It is noteworthy that they also displayed a histogram of $T_{\textrm{eff}}$, which clearly shows the presence (and position) of the gap. The histogram analysis was also demonstrated in the study by D'Antona et al., (2002).

There are clear disadvantages when using histograms for the analysis of data, especially when searching for narrow structures, such as the B{\"o}hm-Vitense gap. The resulting plot (and the results of an automatic analysis of such a plot) will be affected by the choice of several parameters -- the number of bins and the range of binning. The problem can be partially avoided by varying the size of the bins or by switching to the kernel density estimator.

\subsection*{3.3. \textit{k}-nearest neighbours}

The $k$-nearest neighbours ($k$-NN) algorithm is a handy tool for data classification and density estimation. For our work, we are interested in a situation when $k(n)$ is a function of the sample size $n$. We use the rule $k(n)=\sqrt{n}$. As mentioned in the literature (e.g. see Devroye et al., (1996)), this simple option may not always be the ideal approach for data classification. However, we would like to point out that our goal is not classification but to use $k$-NN as a density estimator. In such a case, our choice of $k(n)$ should yield reasonable results, at least in the first approximation.

A rudimentary Python procedure was written to deal with the task. For all data points (Blue Photometer - Red Photometer magnitudes from Gaia) ($BP-RP$, $M_G$)$_i$, an ellipse with a fixed axis ratio $b:a=0.3$ is centred at a given point ($b$ being defined along the $M_G$ axis, $a$ along the $BP-RP$ axis, the ellipse is not rotated). The starting ellipse size is $a=0.001$~mag, which gradually increases with a scaling parameter $s$, until the number of included data points (excluding $i$) reaches $k$. The ellipse scale is then our density estimator and can be plotted against $BP-RP$ and $M_G$. Assuming a uniform (or a nearly uniform) data distribution and starting from the blue end of the CMD, $s$ will sharply decrease towards the data set's median and increase again towards the red end of the CMD. A sudden jump in $s$ is expected if:
\begin{itemize}
\item Point $i$ is an outlier
\item Point $i$ lies in an over-density region
\item Point $i$ lies in an under-density region
\end{itemize}
Over-densities are not expected to be observed in CMDs and can only be the result of statistics. Outliers in the data are most likely to produce only very sharp peaks in the plots of $s$, while under-densities produced by gaps (true or statistical) will be broader and easily recognisable. For this reason, the ellipse scale $s$ is expected to be a reasonable tool for detecting gaps in the CMDs.

In principle, the usefulness of $k$-NN as a gap-detecting technique has already been demonstrated in Ratzenb{\"o}ck et al., (2023). The authors show that a multi-model distribution can be recognised with the help of $k$-NN, with the algorithm used as a density estimator and during their modality test when identifying stellar populations. The $k$-NN algorithm is reasonably accurate (although not necessarily the best) when applied to a task that involves the identification of gaps in data (in our case, in CMDs).

Like the other methods mentioned above, $k$-NN will be affected by statistical gaps in the data. It is unlikely that these gaps can be identified with $k$-NN alone. However, we expect this approach to outperform the others when applied to multiple star clusters.

\subsection*{3.4. \texttt{UniDip} algorithm}

\texttt{UniDip} algorithm is a robust tool used for 1-dimensional data analysis, available as a Python package. It can recognise multi-modal distribution in the data and is based on Hartigan's dip test.

We make use of this algorithm by projecting the data in the CMD into three planes:
\begin{itemize}
\item The $BP-RP$ axis
\item The $M_G$ axis
\item The axis determined by the principal component analysis, see Hotelling, (1936). 
\end{itemize}
The projected data can be analysed using the $UniDip$ package that extracts the regions corresponding to the recognised modes. If a gap exists, it will lie between two neighbouring modes.

One of the advantages of this tool is the ability to use it efficiently for noisy data. Furthermore, \texttt{UniDip} can precisely determine the position of the gaps (unlike in the case of $k$-NN). However, if it is unable to detect a gap in a cluster, the photometric data of that cluster cannot be used in the statistical analysis of the results based on all clusters (no information about a gap is provided).

\subsection*{3.5. Identifying statistical gaps}

One of the main problems with the B{\"o}hm-Vitense gap (or any other true gap) is the ability to distinguish between this gap and a statistical gap that results from poor number statistics. The size of a statistical gap, determined as the difference between two neighbouring points (e.g. in photometric colour), depends heavily on the sample size -- larger samples have smaller sizes than a typical statistical gap.

It is highly non-trivial to determine whether a gap is statistical or a result of a physical phenomenon (or possibly a bias). This is also the main problem why gaps in CMDs remain largely unexplored -- they are often believed to be a result of poor number statistics. This remains problematic, especially when searching for gaps in a large sample of clusters. Removing problematic stars is usually very difficult (automatic approach) and possibly very time-consuming (manual approach). As an alternative, searching for gaps in noisy data of multiple clusters should be possible.

Significant noise in a data set will likely result in a hidden gap. There are two possibilities to distinguish between statistical and actual gaps for a set of multiple clusters. Firstly, one can use the ADM plot and statistically analyse the distances between the points projected onto a given axis. Compared with randomly generated distributions based on the empirical CDF of these values, one can determine the likelihood of finding a gap of a given size in the given sample. Unfortunately, we expect the CMDs to be quite noisy and affected by outliers, making it very unlikely to find an actual gap using this method.

The second option is to use the region in the CMD corresponding to the location of an estimated actual gap (based on one of the above-mentioned algorithms for gap detection). Next, the positions of these regions based on multiple clusters can be stacked together -- the precise approach will depend on the detection algorithm. The statistical gaps should be randomly (not necessarily uniformly) distributed in a given range of values (e.g. photometric colour). If a gap repeats too often, this will point to the possibility of an actual gap. These ranges of values can then be easily checked manually to verify the nature of the gap.

We make use of both approaches for detecting gaps. The second, ``stacking'', method (s-method), together with a manual check, yields a set of gaps which are unlikely of a statistical origin. The first, ``generating'', method (g-method), is used to compare the sizes of the statistical gaps in a given cluster with the estimated size of a given actual gap (based on the s-method). We only use clusters where, in the density estimation, we consider only $N \geq$ 20. 

\section*{4. Different effects for searching the gaps in the CMDs}

We focus on two effects when searching for gaps in the colour-magnitude diagrams of open clusters. Firstly, we look for the classical B{\"o}hm-Vitense gap near the effective temperature of around 6400 K. Our members are taken from Cantat-Gaudin \& Anders, (2020). This effect is explained by changes in the opacity when switching from envelopes dominated by convection to those dominated by radiation (for a more detailed explanation, see de Bruijne et al., (2001). This can be seen in the colour-magnitude diagram as a mini-turnoff from the main sequence, leading to a gap in the data point distribution. As such, we are interested in searching for a gap and aim to detect the mini-turnoffs. These can be detected by manually searching the diagrams and using different photometric colours (combining Gaia and 2MASS photometry). The diagrams are plotted in a region within which the gap is supposed to be located. To be more objective when detecting the mini-turnoff, we also plot the corresponding isochrone - if a significant number of points is located above the curve (and the points in both directions from the mini-turnoff lie around the curve), we can say that a mini-turnoff is likely to be present.

Secondly, we would like to search for other gaps. Several gaps were already detected in the past (for example, those located near 5000 K, 5600 K, 7000 K, and 8000 K; see de Bruijne et al., (2001), Kovtyukh et al., (2004). We are searching for those gaps in colour-magnitude diagrams by counting the number of points in an ellipse centred on a point located on the corresponding isochrone curve (using parameters from Dias et al., (2021). A result is several points as a function of the photometric colour (e.g. $BP-RP$). These numbers can be combined for multiple clusters, with a possibility to introduce metallicity and age filters, and statistically analysed (if a gap is present in a given colour range, we should see a statistically significant decrease in the number of points).

\section*{5. Analysing the Gaps}

We analyse the gaps based on two approaches and discuss them in more detail. 

\begin{itemize}
\item M1: Using ellipses, we can scan the data and count the number of points within the ellipses, then look at how this number changes as a function of photometric colour.
\item M2: There is an older method, which is to sort the array of colour values, then calculate the differences between the neighbouring values (let us call these ``natural gaps''), so we will get something like a cumulative distribution function (not really, but it is an easy transition between this and an empirical CDF). Then, we could use the empirical CDF to simulate data some number of times, always simulating the same number of points as in the observed diagram. Then, one can look at the distribution of the simulated natural gaps, calculate some quantile and see how many natural gaps from the observed cluster are more significant than the value of this quantile (e.g. quantile = 0.95).
\end{itemize}

There are some issues with both approaches. 

\begin{itemize}
\item For M1: The results depend on ``a'' and ``b'' of the ellipse and ``f'' which is the ellipse orientation angle, and there is no single set of values for all clusters, so one has to guess basically. We tried some options for estimating reasonable values automatically but with no success. This means that even if we detect a gap using this method, it will not always work, and therefore, it is useless from a statistical point of view.
\item For M2: This method should work nicely from a statistical point of view, but currently, we can only work with these natural gaps, which are not always helpful. The point is, even when we see that B{\"o}hm-Vitense gap in the diagram, it may be so small that it is statistically the same size as a natural gap (so a statistical gap), there tends to be quite obvious significant gaps which are difficult to deal with using this method.
\end{itemize}

\begin{figure}
    \centering
    \includegraphics[width=8.cm] 
    {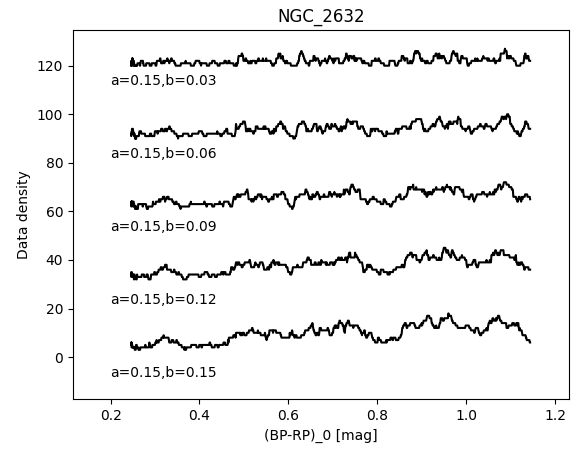}
    \centering
    \includegraphics[width=8.cm]
    {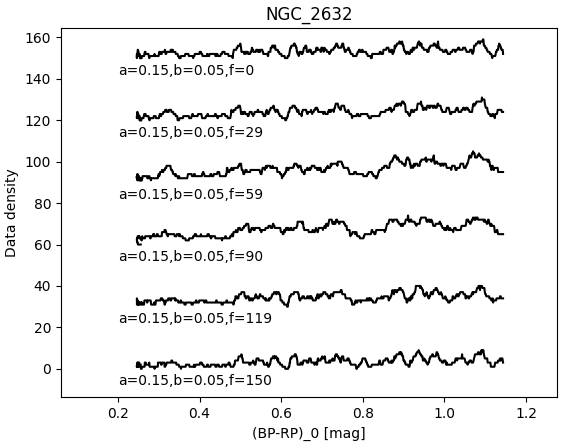}
    \centering
    \includegraphics[width=8.cm]
    {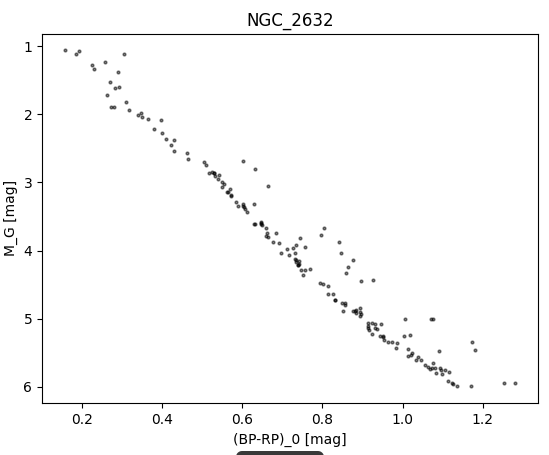}
    \caption{Different ellipses (with the semi-major and semi-minor axis a and b; and f being the orientation angle)
    Panel 1: a is fixed and b is variable. Panel 2: a and b are fixed and f is variable. Panel 3: plot of them in the Color-Magnitude Diagram.}
    \label{fig1}
\end{figure}

Figure \ref{fig1} is based on M1; we tried changing the size of the ellipses (they move along the isochrone that fits the data; we checked, it fits pretty well) and also their orientations, the $B-V$ gap should be somewhere between 0.47 and 0.50 mag, we do not see it in the density diagrams.

\begin{figure}
    \centering
    \includegraphics[width=6.6cm] 
    {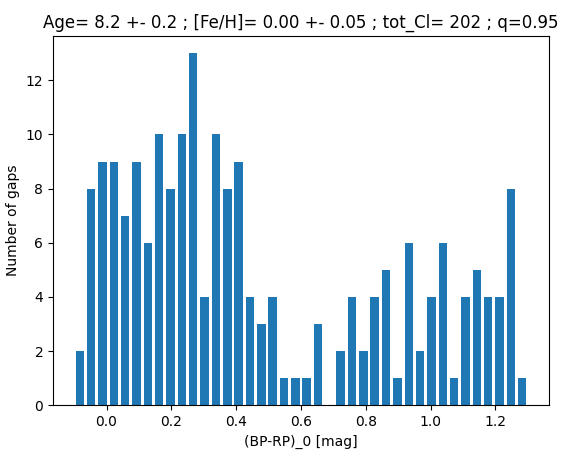}
    \centering
    \includegraphics[width=6.6cm]
    {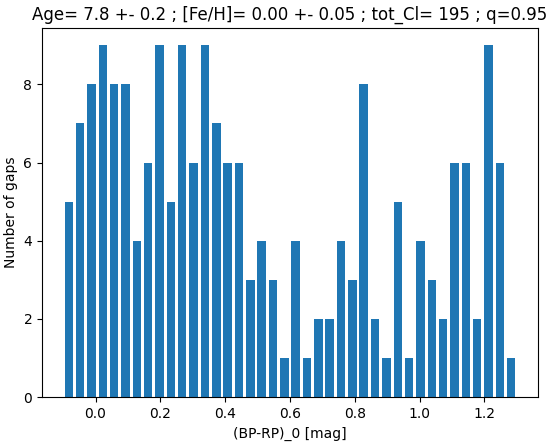}
    \centering
    \includegraphics[width=6.6cm]
    {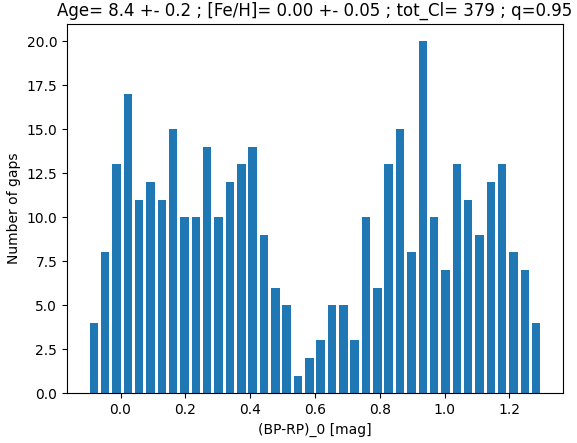}
    \centering
    \includegraphics[width=6.6cm] 
    {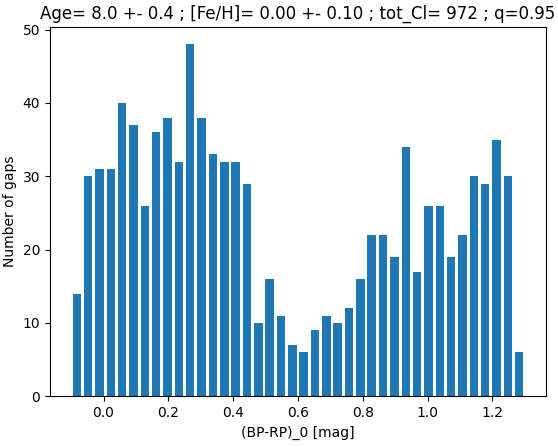}
    \caption{Searching for the B{\"o}hm-Vitense gaps using M2, q=0.95, and the Gaia photometric system.
    Panel 1 to 3: Searching for gaps in different Ages (tot$\_$cl being the number of all investigated clusters), but the same Metallicity and quantile. Panel 4: Searching for gaps with different Age, Metallicity, and tot$\_$cl, but the same quantile as Panel 1 to 3.
    }
    \label{fig2}
\end{figure}

\begin{figure}
    \centering
    \includegraphics[width=6.7cm] 
    {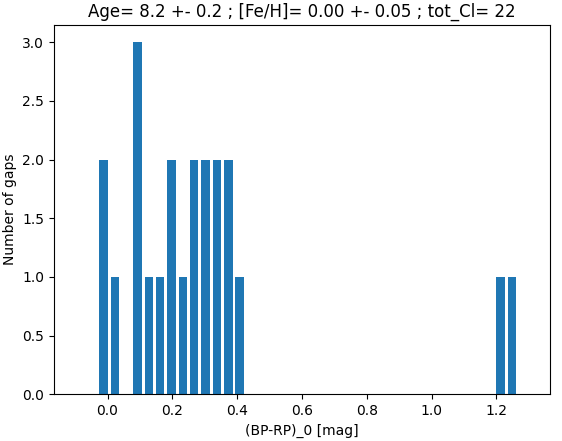}
    \centering
    \includegraphics[width=6.7cm]
    {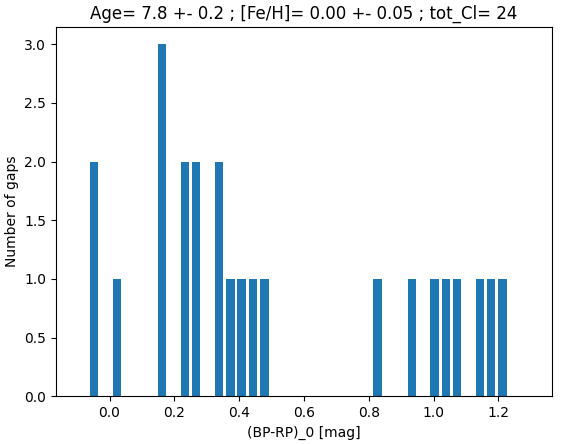}
    \centering
    \includegraphics[width=6.7cm]
    {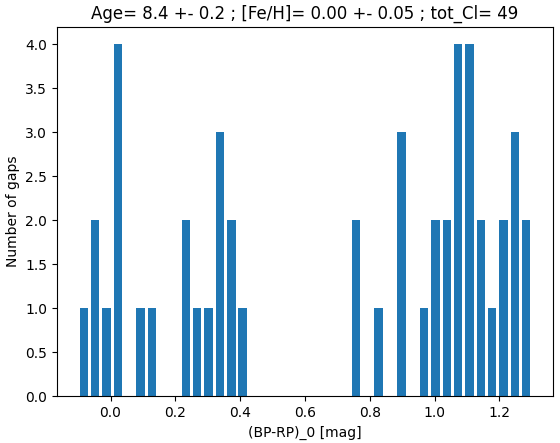}
    \centering
    \includegraphics[width=6.7cm] 
    {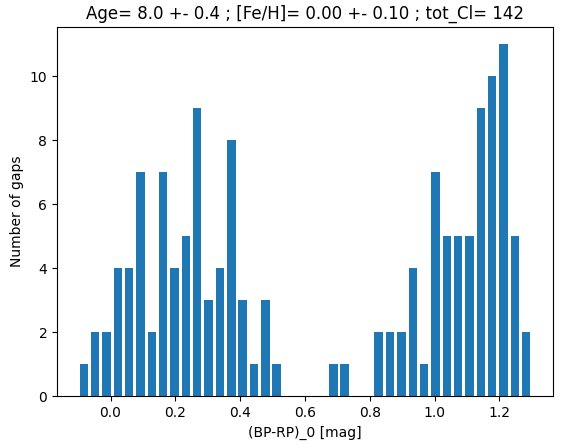}
    \caption{Searching for the B{\"o}hm-Vitense gaps using M2, q=0.99, and the Gaia photometric system.
    Panel 1 to 3: Searching for gaps in different Ages and tot$\_$cl, but the same Metallicity. Panel 4: Searching for gaps with different Age, Metallicity, and tot$\_$cl of the Panel 1 to 3.}
    \label{fig3}
\end{figure}

Figures \ref{fig2} and \ref{fig3} are based on M2; we tried using $q=0.95$ and $q=0.99$. It seems that maybe something shows at around 0.5 mag for the case when $q=0.95$, but it is quite clear that there are other ``gaps'' which are more numerous. 

Overall, neither of the two approaches will be helpful in automatically detecting the gaps. The main problem for M1 is variations in data density at various colour ranges. For M2, the main problem is that the $B-V$ gaps will have a statistically significant different size for most clusters. The ellipses approach is quite helpful in characterizing the gaps in general.

\begin{figure}
    \centering
    \includegraphics[width=7.5cm] 
    {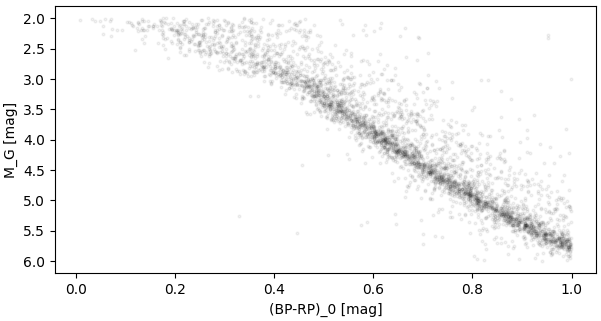}
    \centering
    \includegraphics[width=7.5cm]
    {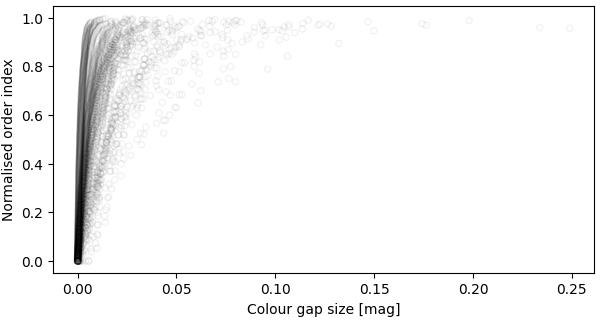}
    \centering
    \includegraphics[width=7.5cm]
    {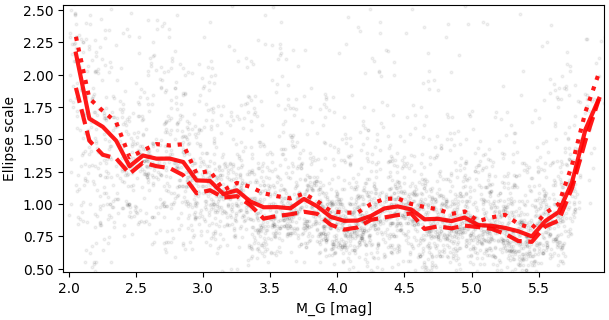}
    \centering
    \includegraphics[width=7.5cm] 
    {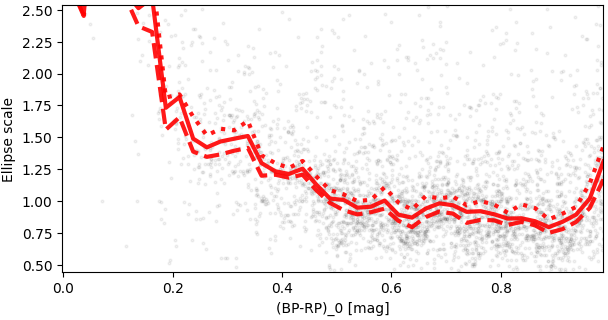}
    \caption{Results for the ellipse scales, BP-RP, and absolute G for the used cluster members, also normalised ellipse scales which filtered by metallicity of [Fe/H] lower than $-$0.02687.}
    \label{fig4}
\end{figure}

\begin{figure}
    \centering
    \includegraphics[width=7.5cm] 
    {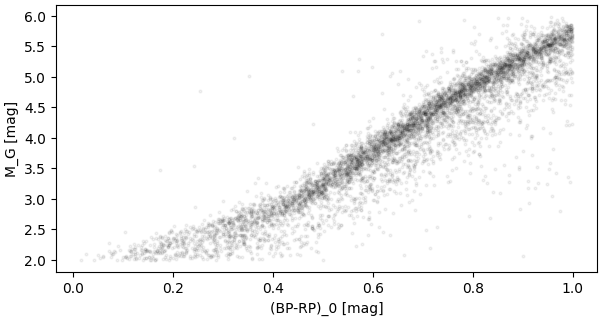}
    \centering
    \includegraphics[width=7.5cm]
    {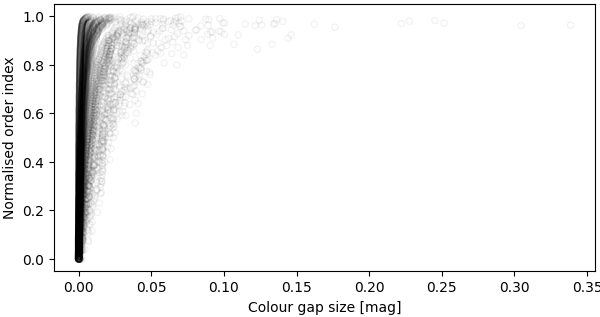}
    \centering
    \includegraphics[width=7.5cm]
    {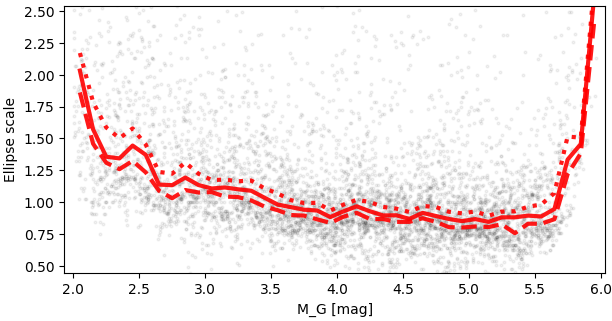}
    \centering
    \includegraphics[width=7.5cm] 
    {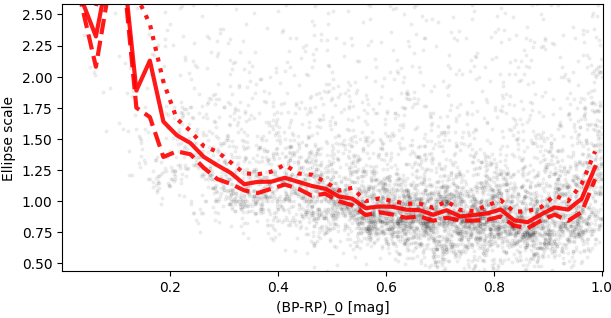}
    \caption{Results for the ellipse scales, BP-RP, and absolute G for the used cluster members, also normalised ellipse scales which filtered by metallicity of [Fe/H] $-$0.02687 to +0.04243}
    \label{fig5}
\end{figure}

\begin{figure}
    \centering
    \includegraphics[width=7.5cm] 
    {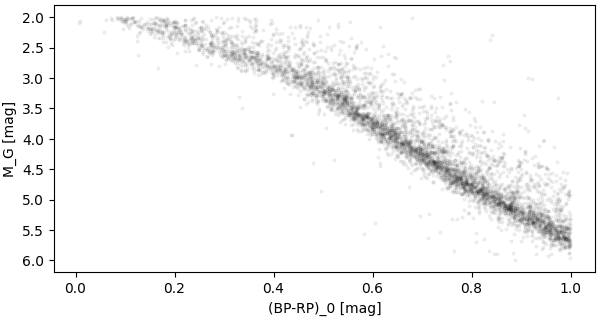}
    \centering
    \includegraphics[width=7.5cm]
    {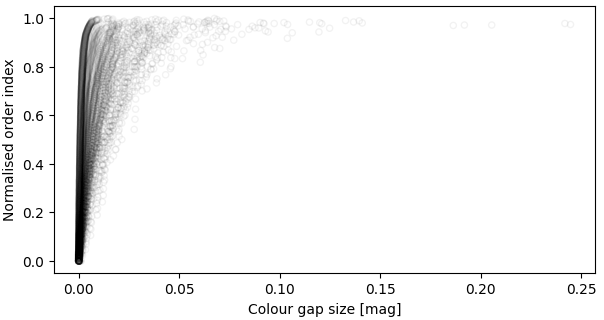}
    \centering
    \includegraphics[width=7.5cm]
    {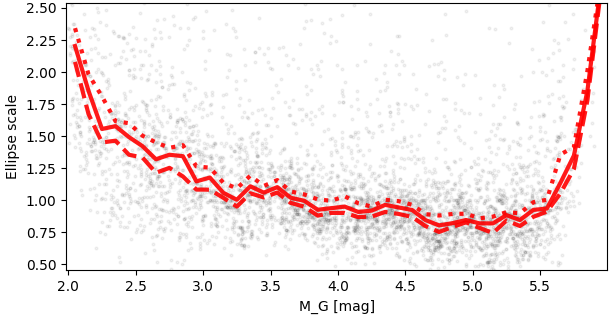}
    \centering
    \includegraphics[width=7.5cm] 
    {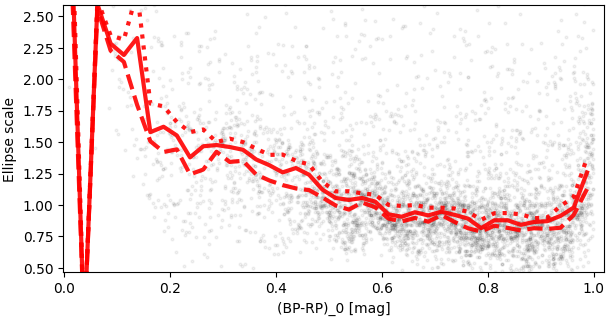}
    \caption{Results for the ellipse scales, $BP-RP$, and absolute $G$ for the used cluster members, also normalised ellipse scales which filtered by metallicity of [Fe/H] higher than +0.04243}
    \label{fig6}
\end{figure}

\begin{figure}
    \centering
    \includegraphics[width=7.5cm] 
    {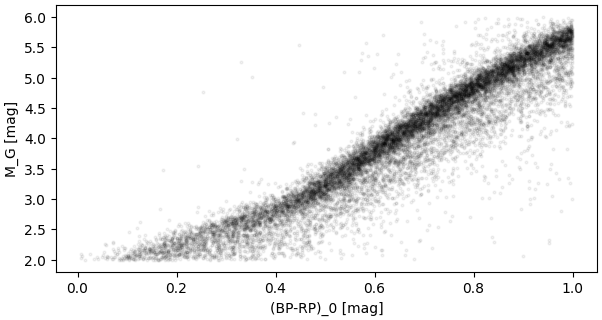}
    \centering
    \includegraphics[width=7.5cm]
    {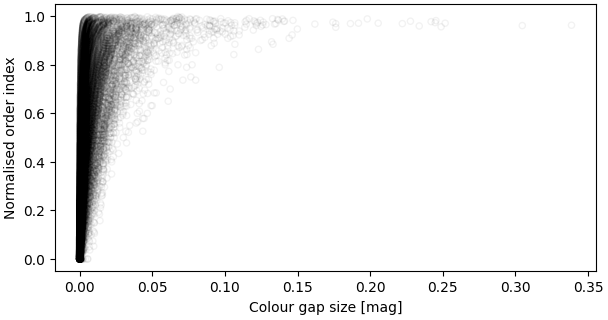}
    \centering
    \includegraphics[width=7.5cm]
    {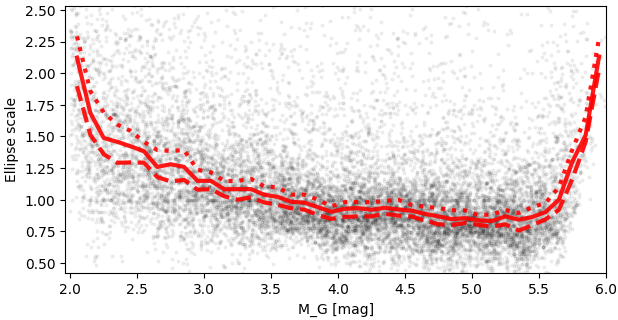}
    \centering
    \includegraphics[width=7.5cm] 
    {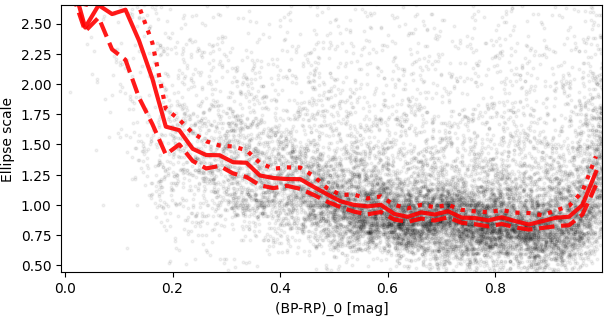}
    \caption{Results for a metallicity of zero with poor number statistics to avoid the errors in cluster parameters in ellipse approach.}
    \label{fig7}
\end{figure}

We used the ellipses approach; in the data files, we found the ranges for the two possible gaps (0.40 to 0.47 and 0.56 to 0.60 mag). Our data files contain ellipse scales, $BP-RP$, and absolute $G$ for the used cluster members ($D \leq 2$ kpc, $A_V \leq 0.5$ mag, $\log t \leq 9.2$, also there must be at least 20 data points for colours range 0 to 1 mag and for $G$ from 2 to 6 mag). For the ``summarized'' figures where all clusters are included, we made use of the normalised ellipse scales (normalised using medians of the scale in the original file, but we also included the file with the normalised values) and filtered by metallicities, which is shown in Figures \ref{fig4} to \ref{fig6}. For the metallicities, we used the following ranges:
\begin{itemize}
\item Z1 = [Fe/H] lower than $-$0.02687
\item Z2 = $-$0.02687 to +0.04243
\item Z3 = higher than +0.04243
\end{itemize}
with the ellipse approach, we are combining multiple clusters, so the errors in cluster parameters may affect the results, but at least we are avoiding the problem with poor number statistics, which is shown in Figure \ref{fig7}.

\section*{Conclusion}
The best option for finding the  B{\"o}hm-Vitense gaps is manually doing it with the help of CMDs. We looked for the gaps for 130 star clusters in the $BP-RP$ ranges between 0.40 and 0.47\,mag, and 0.56 to 0.60\,mag, respectively. We did a statistical analysis of all clusters, detected where the gaps could be found, and identified them in individual clusters. We assumed that the gap is non-statistical if the expected number of such gaps is between 1.0e-6 and 0.5. We saved the location of the gaps and took the five most significant gaps. Then we used the kernel density estimator to see how the gaps are distributed. Finally, the presence of the gaps is being verified manually.

\begin{figure}
    \centering
    \includegraphics[width=7.5cm] 
    {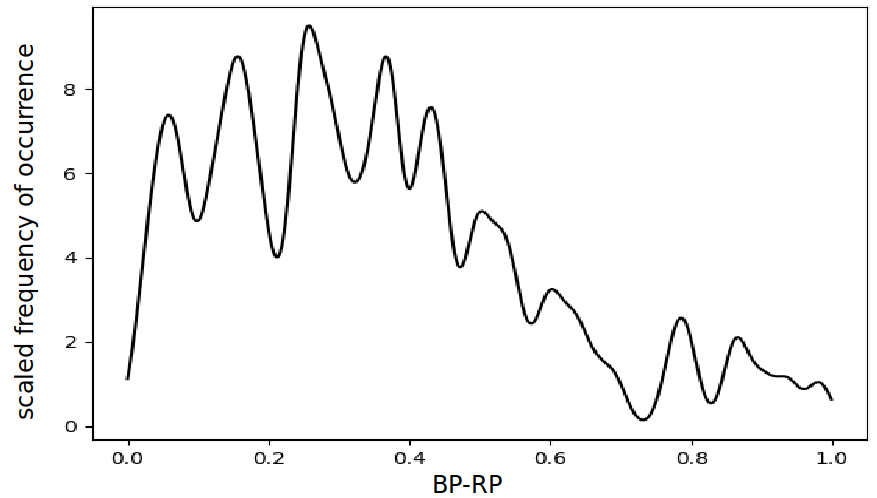}
    \caption{Location of the five largest gaps, between 1.0e-6 and 0.5, respectively, and the primary B{\"o}hm-Vitense gap is barely visible in the gap-frequency diagram.}
    \label{fig8}
\end{figure}

In Fig. \ref{fig8}, the X-axis is intrinsic $BP-RP$, and the Y-axis is scaled frequency of occurrence. We concentrated on the region with colours between 0 and 1\,mag, and we only considered those with at least 100 points in that region. Multiple gaps exist, and the primary B{\"o}hm-Vitense gap is barely visible in the gap-frequency diagram.

\section*{Acknowledgements}

This work was supported by the grant GA{\v C}R 23-07605S and made use of the SIMBAD database, operated at CDS, Strasbourg, France and 
of the WEBDA database, operated at the Department of Theoretical
Physics and Astrophysics of the Masaryk University.

%\newpage


\begin{thebibliography}{}



\bibitem[]{}
%de Bruijne, J.~H.~J. , Hoogerwerf, R., de Zeeuw, P.~T., 2001, {\em  Astron. Astrophys.}, 367, 111

%in the text: de Bruijne et al. (2001)





\bibitem[]{}
Aizenman, M.~L., Demarque, P., Miller, R.~H., 1969, {\em ApJ}, 155, 973






\bibitem[]{}
B{\"o}hm-Vitense, E., 1970, {\em A\&A}, 8, 299





\bibitem[]{} 
B{\"o}hm-Vitense, E., Canterna, R., 1974, {\em ApJ}, 194, 629





\bibitem[]{}
B{\"o}hm-Vitense, E., 1982, {\em ApJ}, 255, 191





\bibitem[]{}
Cantat-Gaudin, T., Anders, F., 2020, {\em A\&A}, 633, A99






\bibitem[]{}
Canterna, R., Perry, C. L., Crawford, D.~L., 1979, {\em PASP}, 91, 263






\bibitem[]{}
D'Antona, F., Montalbán, J., Kupka, F., Heiter, U., 2002, {\em ApJ}, 564, L93






\bibitem[]{}
de Bruijne, J.~H.~J., Hoogerwerf, R., de Zeeuw, P.~T., 2001, {\em A\&A}, 367, 111






\bibitem[]{}
Devroye, L., Lugosi, G., {Gy{\"o}rfi}, L., 1996, {\em Springer}






\bibitem[]{}
Dias, W.~S., Monteiro, H., Moitinho, A., Lépine, J. R. D., Carraro, G., Paunzen, E., Alessi, B., Villela, L., 2021, {\em MNRAS}, 504, 356




\bibitem[]{}
Harris, W.~M., Clarke, J.~T., Caldwell, J.~W., Feldman, P.~D., Bush, B.~C., Cotton, D.~M., Chakrabarti, S., 1993, {\em Optical Engineering}, 32, 30





\bibitem[]{}
Hotelling, H., 1936, {\em Biometrika Trust}, 28, 321






\bibitem[]{}
Jasniewicz, G., 1984, {\em A\&A}, 141, 116







\bibitem[]{}
Kjeldsen, H., Frandsen, S., 1991, {\em A\&A}, 87, 119





\bibitem[]{}
Kovtyukh, V.~V., Soubiran, C., Belik, S.~I., 2004, {\em A\&A}, 427, 933






\bibitem[]{}
Magic, Z., Weiss, A., Asplund, M., 2015, {\em A\&A}, 573, A89





\bibitem[]{}
Mendoza, V., Eugenio, E., 1956, {\em ApJ}, 123, 54






\bibitem[]{}
Rachford, L., Canterna, R., 2000, {\em AJ}, 119, 1296






\bibitem[]{}
Ratzenb{\"o}ck, S., Gro{\ss}schedl, J.~E., M{\"o}ller, T., Alves, J., Bomze, I., Meingast, S., 2023, {\em A\&A}, 677, A59







\bibitem[]{}
Simon, T., Landsman, W.~B., 1997, {\em ApJ}, 484, 360





\bibitem[]{}
Sonoi, T., Ludwig, H.-G., Dupret, M. -A., Montalb{\'a}n, J., Samadi, R., Belkacem, K., Caffau, E., Goupil, M.-J., 2019, {\em A\&A}, 621, A84






\end{thebibliography}
\end{document}